\documentclass[5p,times,twocolumn]{elsarticle}
\usepackage{lineno,hyperref}
\usepackage{graphics,subfigure}
\usepackage{tikz}
\usepackage{pgfplotstable}
\usepackage{pgfplots}
\usepackage{multirow}
\usepackage{float}
\usepackage{textcomp}
\usepackage{amsmath,amsfonts,amssymb,bbm}	
\usepackage{multicol}
\usepackage{blindtext}

\usepackage[utf8]{inputenc}
\DeclareUnicodeCharacter{2212}{-}
\usepackage{booktabs}
\usepackage{tikz}

\newcommand{\squeezeup}{\vspace{-2.5mm}} 

\usepackage[section]{placeins}

\modulolinenumbers[5]

\biboptions{authoryear}

\begin{document}

\begin{frontmatter}

\title{MICDIR: Multi-scale Inverse-consistent Deformable Image Registration using UNetMSS with Self-Constructing Graph Latent}

\author[1,2,3,4]{Soumick Chatterjee\corref{mycorrespondingauthor}}
\cortext[mycorrespondingauthor]{Corresponding author:}
\ead{soumick.chatterjee@ovgu.de}

\author[1]{Himanshi Bajaj}
\author[1]{Istiyak H. Siddiquee}
\author[1]{Nandish Bandi Subbarayappa}
\author[1]{Steve Simon}
\author[1]{Suraj Bangalore Shashidhar}
\author[3,5,6]{Oliver~Speck}
\author[1,2,6]{Andreas N{\"u}rnberger}

\address[1]{Faculty of Computer Science, Otto von Guericke University Magdeburg, Germany}
\address[2]{Data and Knowledge Engineering Group, Otto von Guericke University Magdeburg, Germany}
\address[3]{Biomedical Magnetic Resonance, Otto von Guericke University Magdeburg, Germany}
\address[4]{Genomics Research Centre, Human Technopole, Milan, Italy}
\address[5]{German Centre for Neurodegenerative Disease, Magdeburg, Germany}
\address[6]{Centre for Behavioural Brain Sciences, Magdeburg, Germany}

\begin{abstract}
Image registration is the process of bringing different images into a common coordinate system - a technique widely used in various applications of computer vision, such as remote sensing, image retrieval, and, most commonly, medical imaging. Deep learning based techniques have been applied successfully to tackle various complex medical image processing problems, including medical image registration. Over the years, several image registration techniques have been proposed using deep learning. Deformable image registration techniques such as Voxelmorph have been successful in capturing finer changes and providing smoother deformations. However, Voxelmorph, as well as ICNet and FIRE, do not explicitly encode global dependencies (i.e. the overall anatomical view of the supplied image) and, therefore, can not track large deformations. In order to tackle the aforementioned problems, this paper extends the Voxelmorph approach in three different ways. To improve the performance in case of small as well as large deformations, supervision of the model at different resolutions has been integrated using a multi-scale UNet. To support the network to learn and encode the minute structural co-relations of the given image-pairs, a self-constructing graph network (SCGNet) has been used as the latent of the multi-scale UNet - which can improve the learning process of the model and help the model to generalise better. And finally, to make the deformations inverse-consistent, cycle consistency loss has been employed. On the task of registration of brain MRIs, the proposed method achieved significant improvements over ANTs and VoxelMorph, obtaining a Dice score of 0.8013±0.0243 for intramodal and 0.6211±0.0309 for intermodal, while VoxelMorph achieved 0.7747±0.0260 and 0.6071±0.0510, respectively.

\end{abstract}


\end{frontmatter}


\section{Introduction}
\label{ch:introduction}
Image registration is the process where two given images are aligned to an identical geometrical coordinate system~\citep{boveiri_medical_2020}. This process has been widely used in satellite imagery, computer vision, military applications, etc. One of these fields is medical imaging~\citep{de_vos_deep_2018}. The clinical importance of image registration in the field of medical imaging cannot be emphasised enough. As a critical component of modern medical practice, image registration serves to align and integrate multiple images, often from disparate imaging modalities or acquired at varying time points, into a unified, coherent representation~\citep{hill2001medical}. It enables clinicians to accurately assess and monitor disease progression, evaluate treatment response, and plan targeted interventions~\citep{crum2004non}. Furthermore, the fusion of multimodal data facilitates the extraction of valuable quantitative information, ultimately leading to enhanced diagnostic precision, improved treatment planning, and, most importantly, better treatment outcomes~\citep{pluim2003mutual}. Consequently, the continued advancement of image registration methodologies not only promotes the optimisation of clinical workflows but also represents a vital cornerstone in the quest for personalised medicine and precision healthcare~\citep{klein2007evaluation}.

Medical image registration tasks can be divided into two categories: intramodal - registering images of the same modality (e.g. MRI-T1 or MRI-T2 or CT) of different subjects into a standard coordinate system, and intermodal - registering images of different modalities (e.g. MRI-T1 to MRI-T2, or MRI to CT) of the same subject. In image registration terminology, the image which will be registered or "moved" to the coordinate of a second image is known as the moving image (or source image), while the second image is known as the fixed image (or destination image). There are various ways to perform the image registration task. Traditional approaches focus on directly optimising pairs of images based on some criteria or function. More modern approaches include machine learning and deep learning based solutions that try to understand the problem as a whole rather than focusing on individual pairs of images. Such approaches have far-reaching advantages, e.g. they can be very fast, and they can have the capability of accommodating a wide array of deformations. Furthermore, some of those solutions do not even require manually delineated training data, meaning that some of those algorithms are unsupervised in nature, thus reducing the human effort and possibility of human errors in data preparation. 

Traditional approaches for performing medical image registration usually solve some optimisation function for each image pair. This optimisation function is often defined using a similarity metric that allows the non-linear mapping of apparently similar voxels while maintaining local smoothness, etc. constraints~\citep{zhang_inverse-consistent_2018}. This is commonly known as a non-learning-based solution. Because such solutions require optimising the same criteria from scratch for each image pair, it requires an extensive amount of time and computational effort to solve such an optimisation problem. To speed up this process, supervised machine/deep learning methods can be used. Here, the optimisation does not have to take place for every pair of images every time. Instead, the learnt parameters of the registration model can be used to transform a newly obtained image. But such methods require a considerable amount of manually delineated training data that is difficult to obtain and, most importantly, more vulnerable to human error. 

Deep learning based solutions try to map the moving images to fixed images by learning a generalisable representation of the displacement vector field that represents the change in the two images. Both the images are fed into the network, and the network attempts to learn the parameters to best transform the image from one to another. Plenty of deep learning based solutions have been proposed, and most of these solutions use unsupervised learning~\citep{de_vos_deep_2018,balakrishnan_voxelmorph_2019,kim_cyclemorph_2020}. Some of these approaches are based on Generative Adversarial Networks (GAN)~\citep{lin_st-gan_2018,mahapatra_gan_2019} - which is a class of self-criticising neural networks. But, owing to the fact that GANs are prone to generate unrealistic imagery and the difficulty of explaining them might not always be a preferred solution. However, there are some intuitive yet efficient ways of regularising these GAN-based networks for producing realistic images, such as cycle consistency \citep{zhang_inverse-consistent_2018,kim_cyclemorph_2020}, bio-mechanics informed regulariser \citep{qin_biomechanics-informed_2020}, etc. However, the existing research either avoided considering or failed to incorporate information about structural connectivity inside the brain. 

The aim of this research is to develop a solution using unsupervised deep learning that is capable of registering images for both intermodal and intramodal cases, which can better integrate structural information into the solution and is able to handle small as well as large deformations.
\subsection{Related Work}
\label{ch:baselinemethods}
Several deep learning based techniques have been proposed over the years to perform inter- and intramodal image registration. \citet{de_vos_deep_2018} focused on unsupervised learning whereby registration is done by calculating a dense displacement vector field (DVF) between the moving and the fixed image. In order to achieve this DVF, CNNs weights are used as parameters of the DVF and optimised to calculate the transformation parameters. ICNet~\citep{zhang_inverse-consistent_2018} employed inverse consistency and prepared a pipeline in an unsupervised fashion for deformable image registration, where both fixed and moving images are symmetrically deformed toward each other. In order to facilitate the production of realistic images, they introduced an anti-folding constraint and smoothness constraint that restricts the amount of "folds" in the warped image and ensures the smoothness of the produced image, respectively. FIRE~\cite{Wang2019} focuses on unsupervised image registration for intermodality. To achieve this, it uses cycle consistency loss and inverse-consistent property. Voxelmoroph~\citep{balakrishnan_voxelmorph_2019} works with concatenated fixed and moving images and predicts the DVF of the given image pair, which is then applied on the moving image using a spatial transformer. ADMIR~\citep{tang_admiraffine_2020} - Affine and Deformable Medical Image Registration is an end-to-end method for affine and deformable image registration utilising CNNs. This method does not require the images to be pre-aligned, which in turn helps to do image registration quickly with good accuracy.

\subsection{Contributions}
This paper proposes MICDIR - \textbf{M}ulti-scale \textbf{I}nverse-\textbf{C}onsistent \textbf{D}eformable \textbf{I}mage \textbf{R}egistration, which utilises the proposed MSCGUNet - \textbf{M}ulti-scale \textbf{UNet} with \textbf{S}elf-\textbf{C}onstructing \textbf{G}raph Latent, for deformable image registration for 3D volumetric images. The method performs image registration at two different scales (i.e. original resolution, as well as down-scaled resolution) using the multi-scale UNet architecture, tries to encode the semantics to improve generalisation using the self-constructing graph network as the latent space of the said UNet, and attempts to make the method inverse-consistent using the cycle consistency loss. The proposed method has been employed for the task of intramodal and intermodal image registration of brain MRIs and has been compared against two non-deep learning and three deep learning baselines. Finally, the paper shows the result of an ablation study to understand the contributions of the three aforementioned components.   

\section{Methodology}
\label{ch:proposedmethods}
This section explains the proposed methodology - MICDIR, and its nuances. It starts by explaining the different background methods utilised for building the model, then furnishing the details about the proposed method - including the hypotheses behind choosing the different components, the proposed network architecture, training procedure; explains the baseline methods against which the proposed method has been compared, mentions the dataset that has been utilised, and finally elucidates the different evaluation methods. 

\subsection{Background}
\label{ch:background}
UNet, spatial transformers, and the SCGNet are three of the main building blocks of the proposed method - which are explained here.

\subsubsection{UNet}
One of the most popular network models in the field of computer vision is the UNet model~\citep{Ronneberger2015}, which was initially proposed for the task of semantic segmentation. The network consists of a contracting path that generates an image pyramid of features using pooling and convolution operators, whilst the expanding path increases the resolution of the image through upsampling. It uses the output of the contracting path to localise the feature and uses multiple channels combined with convolution to learn about semantics \cite{Ronneberger2015}. 3D UNet \cite{iek20163DUL} extends the vanilla 2D UNet architecture to 3D to work with volumetric data. UNet has been extended as Multi-scale UNet by utilising deep~\citep{Zeng2017,Bortsova2017} or multi-scale supervision~\citep{Zhao2019} - which can improve the training quality of the network with better gradient flow~\citep{Zeng2017} and aids in learning better discriminative features~\citep{Bortsova2017} - shown to significantly improve the performance over the vanilla UNet~\citep{Chatterjee2020DS6DL}.

\subsubsection{Spatial Transformers}
CNNs can successfully accomplish tasks on spatially equivariant data and are limited in being invariant to the input data. Spatial transformers \cite{jaderberg_spatial_2016} help bridge this gap by learning the warping parameters that could make the network invariant whilst being trained along with the same CNN network without extra supervision. The localisation network consists of learnable parameters that learn the warping parameters, such as affine transformation matrix or deformation field, given the input feature maps. The grid generator accepts the parameters from the localisation network and transforms the input. The transformation is conditional to the input and actual pairs. This means the spatial transformer localisation network needs to learn to adjust the affine or deformable parameters necessary to get closer to the actual image.

\subsubsection{Graph Neural Networks and SCGNet}
Graph Neural Networks (GNN)~\cite{Bronstein2017GeometricDL} is a type of neural network that operates on graph-like data structures. Many real-world problems could be modelled in terms of graphs, such as recommender systems, social media user preferences, etc. Images could be crudely modelled as graphs where each pixel is represented as a node and connected with its eight neighbours. Key points extracted from an image could serve as nodes, and their connections are edges. 
 
Transforming an image into a full graph could be difficult due to the number of nodes and edges. A large number of nodes and edges could cause computational issues and would not be a better representation. It would be easier to convert features from an encoder into a graph to reduce the computation and increase the receptive field simultaneously. SCGNet \cite{Liu2020SelfConstructingGC} helps to bridge this gap wherein it can self-construct a graph from a 2D feature map and map it into vertices and edges in latent space. 

The Encoder module uses adaptive pooling to reduce the dimensionality of the feature maps and transform it into a one-dimensional latent space of \(n = \overline(h) * \overline(w) * d\) nodes where \(\overline{h}\) and \(\overline{w}\) are spatial dimensions of the encoder output. This is further transformed into a distribution represented by mean \(\mathcal{M}\) and standard deviation \(\Sigma\).

\begin{equation}
Mean(\mathcal{M}) = Flatten( Conv_{3\times3}( \overline{Feature Maps}))
\end{equation}

\begin{equation}
Standard Deviation(\Sigma) = Log( Conv_{1\times1}( \overline{Feature Maps}))
\end{equation}

The mean and standard deviation are reparameterised using a standard Gaussian noise \(\gamma\mathcal{N}(0,1) \). This helps to centre the distribution during training as shown in equation \ref{eq:mean_std} \cite{Liu2020SelfConstructingGC}.

\begin{equation}
\mathcal{Z} = \mathcal{M} + \Sigma \cdot \gamma
\label{eq:mean_std}
\end{equation}

The Kullback-Leibler divergence loss \cite{Melbourne2010ImageSM} is used to minimise the loss along with the residual loss \(\hat{Z} = \mathcal{M}\cdot(1-log\Sigma)\). The logarithm is applied on standard deviation to make the distribution monotonic and stabilise the training regimen as shown in equation \ref{eq:kl_divergence}

\begin{equation}
    \mathcal{L}_kl = \frac{-1}{2nc} \sum_{i=1}^{n} \sum_{j=1}^{c} (1 + log(\Sigma_{ij})^2  \mathcal{M}^2  (\Sigma_{ij})^2) 
    \label{eq:kl_divergence}
\end{equation}

The encoder produces node information \(\mathcal{Z}\) but not edge details. This is produced by a decoder by generating an Adjacency matrix from nodes as shown in equation \ref{eq:dot_product}. This produces a weighted and undirected graph where similar nodes will have values close to one, and dissimilar nodes will have values close to zeros
\begin{equation}
    A = ReLU( \mathcal{Z}  \cdot \mathcal{Z} ^ T )
    \label{eq:dot_product}
\end{equation}

The node and edge information are further passed to a  GNN network, which uses \( ReLU(Conv_{3\times 3} )\) to produce a graph of size \( \overline{h} * \overline{w} * d \) nodes. 

\subsection{Proposed Method: MICDIR} \label{proposed_method_mscgunet}
Deep learning based deformable image registration techniques such as Voxelmorph \cite{balakrishnan_voxelmorph_2019} have been successful in capturing fine changes and providing smooth deformations. However, Voxelmorph, as well as ICNet \cite{zhang_inverse-consistent_2018} and FIRE \cite{Wang2019} do not explicitly encode global dependencies (e.g. an overall anatomical view of the supplied image in case of medical images) and track large deformations (i.e. large differences between the images). This research tries to address these issues by employing a self-constructing graph network~(SCGNet)~\cite{Liu2020SelfConstructingGC} to encode semantics - which might improve the learning process of the model and help the model to generalise better, multi-scale supervision to be able to work well in case of small as well as large deformations, and cycle consistency (also known as inverse consistency or IC) for making the deformations consistent in both forward (moving-fixed) and inverse (fixed-moving) directions.   

\subsubsection{Hypotheses}
The MSCGUNet tries to capture global dependencies using SCGNet \cite{Liu2020SelfConstructingGC}, wherein encoded CNN features are used to identify semantics. The authors hypothesise that these dependencies could help the model to learn relationships even between two distantly located structures in the fixed and moving image and help generate better deformations. SCGNet can find clusters of similar information in the image using latent space features~\cite{Liu2020SelfConstructingGC}. If the model knows about the landmarks, then it would help improve the deformation field. 

\citet{Chatterjee2020DS6DL} uses multi-scale supervised UNet to perform segmentation of vessels, where the multi-scale supervision has been employed to handle a different amount of deformations, where deformations are predicted, and losses are calculated for both the original image and a downsampled image. The authors hypothesise that deformations on the downsampled image would cover for larger deformations while the deformations on the original image would produce finer deformations. The multi-scale supervision also allows for faster convergence and brings stability to the training as there would be multiple channels through which gradients can flow.

An adapted version of the cycle consistency loss, inspired by \cite{Zhou2016LearningDC}, was employed to maintain consistent deformations. Cycle consistency, also known as inverse consistency (IC), helps to maintain flow correspondence between fixed and moving images, thereby drastically reducing the number of possible deformation fields to achieve the same deformation in both forward (deformation from fixed to moving) and inverse (deformation from moving to fixed) directions. 

\subsubsection{Method Overview} 
\begin{figure*}\centering
    \centering
    \includegraphics[width=0.8\textwidth]{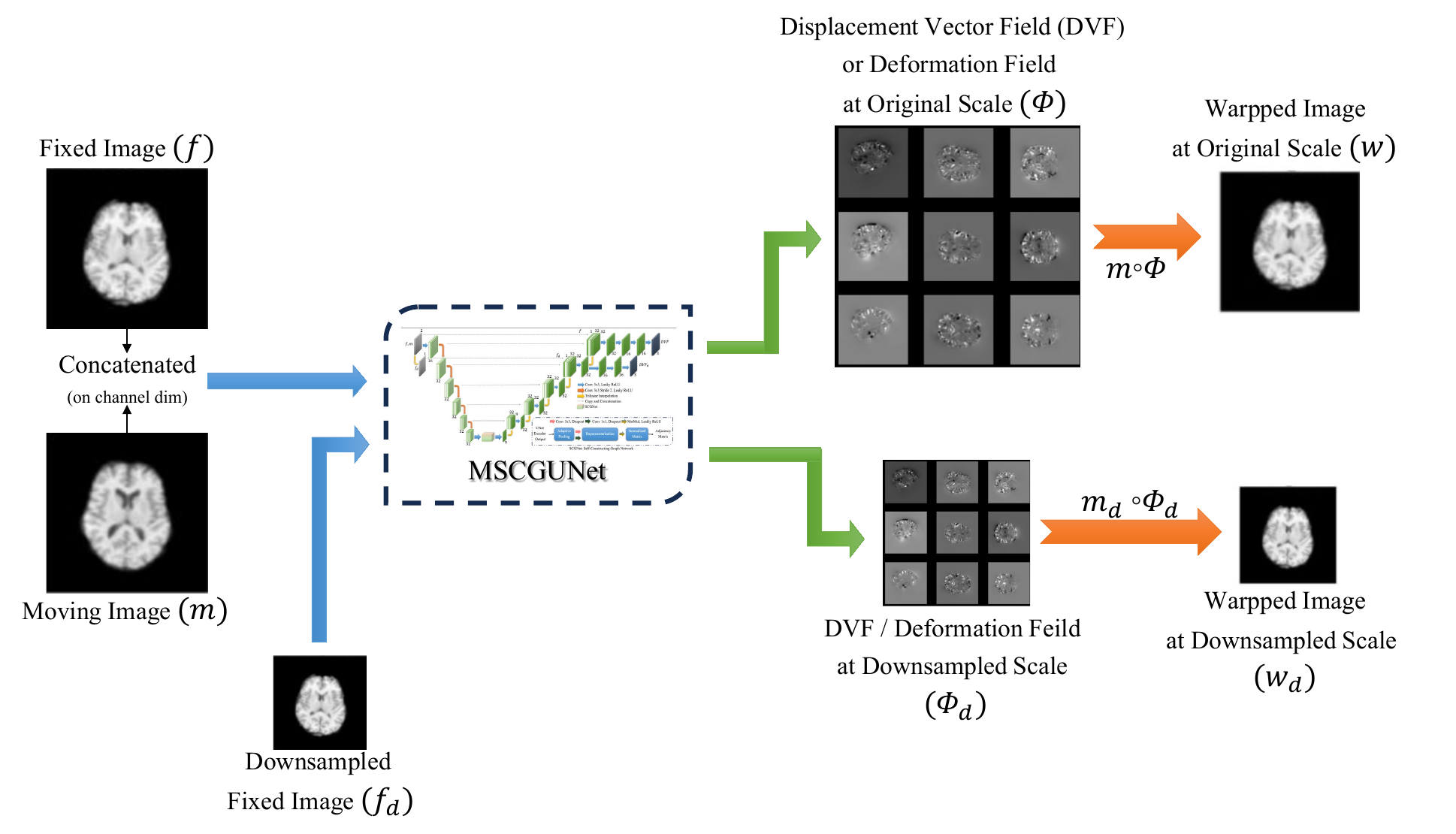}
    \caption{Method Overview: MICDIR - \textbf{M}ulti-scale \textbf{I}nverse-\textbf{C}onsistent \textbf{D}eformable \textbf{I}mage \textbf{R}egistration}
    \label{fig:Multi Scale UNet Overview}
\end{figure*}

Let \(f\), \(m\) be fixed and moving nD images where they belong to domain \( \Omega \subset \mathbb{R}^n\). The network model parameterises the function \(G_\theta\) that calculates the deformation field \(\phi\)  given a pair of fixed \(f\) and moving \(m\) images. The deformation field will have n-channels as it has to represent displacement in each of the n-axes. If it is a 3D image (also known as volume), then the deformation field will have three channels where each channel represents displacement in the x, y and z axes. The proposed network model MSCGUNet takes in the fixed \(f\) and moving \(m\) image along with its downsampled counterparts \(f_d\) and \(m_d\) and computes deformation fields \(\phi\) and \(\phi_d\). Both pairs are passed through the network, and corresponding deformation fields are then applied to \(m \circ \phi \) and \(m_d \circ \phi_d \). Spatial transformers are used to warp the moving image, and a similarity metric is used to compare \(f\) with \(m\) and \(f_d\) with \(m_d\). Gradient descent is used to minimise the loss and find optimal parameters \(\hat{\theta}\). Fig \ref{fig:Multi Scale UNet Overview} provides an overview of the MICDIR method.

\squeezeup

\subsubsection{Architecture of MSCGUNet - the proposed network} 
\begin{figure*}\centering
    \centering
    \includegraphics[width=\textwidth]{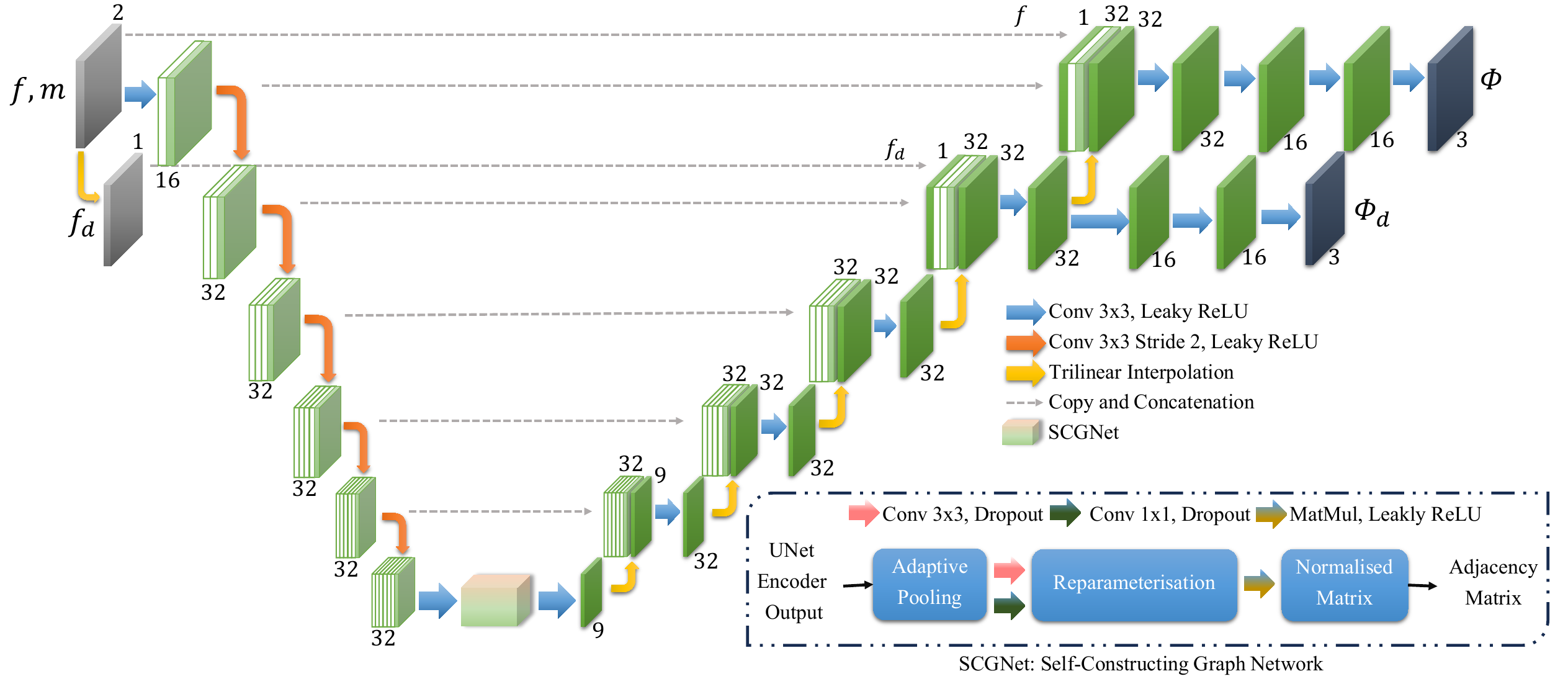}
    \caption{Architecture of MSCGUNet: \textbf{M}ulti-scale \textbf{UNet} with \textbf{S}elf-\textbf{C}onstructing \textbf{G}raph Latent}
    \label{fig:Multi Scale UNet with SCG Architecture}
\end{figure*}

The fixed \(f\) and moving \(m\) images are concatenated on the channel dimension and supplied together to the network. The proposed model uses a UNet structure to encode and decode the image information. For encoding, it uses convolution with a stride of 2 and LeakyReLU activation to downsample the image and learn relevant features. Encoders near the latent space learn more high-level features, whilst encoders at the beginning learn low-level features that act as an image pyramid. The encoded information is fed into the latent space of MSCGUNet, which is a self-constructing graph neural network (SCGNet). The SCGNet first parameterises the encodings to latent embeddings with mean and standard deviation and derives an undirected graph through an inner product of latent embeddings. The output is then fed into a fully-connected graph convolution network (GCN) which attempts to learn the semantics of the image (e.g. brain anatomy in the case of brain MRIs). The GCN output is upsampled through trilinear interpolation and supplied to a CNN decoder, which also takes input from the respective encoder as skip connections. Each decoder block scales up the spatial dimensions of the image and reduces the channels by a factor of \(2\). Skip connections help improve gradient backpropagation and training convergence. It also supplies the corresponding encoder information that provides location information that would be missing from the decoder. Fixed image \(f\) and \(f_d\) are concatenated to the UNet output to maintain the contrast of the fixed image \cite{Dewey2019DeepHarmonyAD}. Finally, deformation fields are smoothed by convolution layers which help learn deformation at finer scales. Fig \ref{fig:Multi Scale UNet with SCG Architecture} illustrates the model architecture.


\squeezeup

\subsubsection{Training Procedure} 
Fixed \(f\) and moving \(m\) images, along with their downsampled counterparts \(f_d\) and moving \(m_d\) are used to train the network. \(f\) and \(m\) are given as input to the MSCGUNet, and it provides deformation fields at both original and downsampled scale, \(\phi\) and \(\phi_d\), respectively. They are then applied on \(m\) and \(m_d\) to obtain the warped images \(w\) and \(w_d\), and are finally compared against \(f\) and \(f_d\) to compute the loss. This process is repeated for the same image pair by swapping fixed and moving images - to obtain the inverse consistency loss. Finally, both losses are added up and backpropagated through the network, shown in Eq.~\ref{eq:total_mscgunet_loss}. 

\begin{equation}
\begin{split}
l_{total} =  \alpha(l_{sim_{f->m}} + l_{sim_{m->f}})  + \alpha_d  (l_{sim_{f_d->m_d}} + l_{sim_{m_d->f_d}})  \\+ 
\beta(l_{sm_{f->m}} + l_{sm_{m->f}})  + \beta_d(l_{sm_{f_d->m_d}} + l_{sm_{m_d->f_d}})  \\+
\lambda( l_{scg_{f->m}} + l_{scg_{m->f}} )  
\end{split}
\label{eq:total_mscgunet_loss}
\end{equation}

Normalised cross-correlation (NCC) and normalised mutual information (NMI) were used as the similarity measure \(l_{sim}\), for intramodal and intermodal registration, respectively. Smoothness loss \(l_{sm}\) \cite{Horn1981DeterminingOF} acts as a regulariser that penalises sharp deformations, and SCG loss \(l_{scg}\) helps to learn similar features in the input image pairs. \(\alpha\) , \(\alpha_d\),  \(\beta\), \(\beta_d\), and \(\lambda\) in the equation are hyper-parameters - coefficients to adjust the impact of each component in the final optimisation process. After experimenting with various values, the final values were set to be \(-1.2\), \(-0.6\), \(0.5\), \(0.25\), and \(5\), respectively. It is noteworthy that this set of hyper-parameters was suitable for the experiment setup (dataset - imaging modalities, preprocessing steps used, etc.) and for other setups -  different sets of hyper-parameters might result in better performance. In this research, the network was trained using 200 volumes for 1000 epochs using the aforementioned loss (Eq.~\ref{eq:total_mscgunet_loss}), which was optimised using the Adam optimiser with a learning rate of \(3e-4\). The code of MICDIR is available on GitHub~\footnote{MICDIR on GitHub: \url{https://github.com/soumickmj/MICDIR}}.

\subsection{Baselines}
The performance of the proposed model was compared against a non-deep learning based baseline, a direct optimisation technique, and three deep learning based baselines. Those baselines and their implementation configurations are explained here.

\subsubsection{Non-Deep Learning Baseline: ANTs}
Advanced Normalisation Tools (ANTS)~\citep{avants2009advanced} is regarded as one of the state-of-the-art medical image registration and segmentation toolkit and is considered to be a gold standard in comparative studies. ANTS, a non-deep learning framework, offers several types of linear and non-linear transforms and different choices of optimisation metrics. The typical optimisation metrics include mutual information, cross-correlation, mattes and demons. For deformable intramodal registration, symmetric normalisation (SyN)~\citep{avants2008symmetric} transform with cross-correlation optimisation metric was used, while for intermodal, the same transform was used with mutual information as optimisation metric.

\subsubsection{Naive Baseline: Direct Optimisation} \label{proposed_method_directoptimization}
The previous sections have explained the technique of parameterising a function using large deep learning models to calculate deformation fields. For the direct optimisation idea, the usage of deep learning models is ditched, and the deformation field is directly optimised using gradient descent algorithms without any model parameters. The rationale behind direct optimisation is to experiment on registration performance without any deep learning model or complex mathematical optimisation procedure. This idea is mathematically supported by \citet{fuse2000comparative}, which describes various techniques based on gradients to estimate optical flow, and \citet{Sherina2020DisplacementFE}, which discusses using multi-scale, iterative, numerical approaches and using additional speckle information to solve registration problems.

In these experiments, an optimiser is directly applied to a deformation field of size \( 3 \times 128 \times 128 \times 128 \) for a pair of images, and then the deformation field is optimised using image similarity metrics such as NCC or NMI and smoothness loss. Equation \ref{eq:direct_optimzation_loss} explains the loss function used to optimise the deformation field. \(f \) and \(m \) represents fixed and moving image, \( l_{sim_{m->f}} \) represents similarity loss and \(l_{sm_{m->f}}\) represents smoothness loss to regularise deformation field. This research used the PyTorch implementations~\footnote{PyTorch Optimisers: \url{https://pytorch.org/docs/stable/optim.html\#algorithms}} of four different optimisers: Adam~\citep{Kingma2014}, AdamW~\citep{loshchilov2017decoupled}, RMSProp~\citep{hinton2012neural,graves2013generating}, and SGD~\citep{robbins1951stochastic,sutskever2013importance}.

\begin{equation}
l_{total} =  \alpha(l_{sim_{m->f}})  + \beta(l_{sm_{m->f}}) 
\label{eq:direct_optimzation_loss}
\end{equation}
-
\subsubsection{Deep Learning Baseline: ICNet}
\citet{zhang_inverse-consistent_2018} performs image registration both ways - moving to fixed and fixed o moving. Let's assume one image is $A$ and another is $B$. with ICNet architecture, image $A$ is registered to image $B$ and called $warped\_A$, and image $B$ registered to image $A$ is called $warped\_B$. When image $A$ gets registered to image $B$, in the process, a flow or displacement vector field $Flow\_AB$ is generated. Similarly, when image $B$ gets registered to image $A$, flow $Flow\_BA$ is generated.   

ICNet \cite{zhang_inverse-consistent_2018} proposed three types of constraints, inverse-consistent constraint, anti-folding constraint, and smoothness constraint. Inverse consistent is a regularisation term used for penalising the deviation of the transformation fields (flow) from their corresponding opposite or inverse mappings, i.e. inverse flow. The second constraint is the anti-folding constraint to prevent folding from happening in the flow generated. The last constraint used is the smoothness constraint. This constraint encourages the local smoothness of transformation fields. For the objective function, the authors used mean squared distance to measure the similarity between the fixed image and the registered image. So, the final objective function becomes the sum of similarity, inverse consistency, anti-folding, and smoothness constraints, where the constraints are multiplied by hyperparameters $a$, $b$, and $c$, respectively, to balance out their values. 

\paragraph{Architecture and Method Details}
ICNet \cite{zhang_inverse-consistent_2018} takes image $A$ and $B$ as input, and results in $warped\_A$ and $warped\_B$ by modelling $Flow\_AB$ and $Flow\_BA$. Authors have used two UNet \cite{Ronneberger2015,iek20163DUL} models for registering, one for image $A$ to $B$, $B$ being the target image and another one for image $B$ to $A$, $A$ is the target image. These two models share parameters and have the same structure. 
The ICNet \cite{zhang_inverse-consistent_2018} pipeline consists of a grid sampler that takes as input image $A$ and flow $Flow\_AB$ and gives $warped\_A$. Similarly, image $B$ and flow $Flow\_BA$ as inputs result in $warped\_B$ as the output. A grid sampling network is nothing but a spatial transformer network \cite{jaderberg_spatial_2016} that is fully differentiable. It consists of a generator that generates a spatial grid and a sampler. This spatial grid gets converted to a sampling grid by using the displacement vector fields (DVF) or flow obtained from each of the UNet models. Finally, the source image is warped using bilinear interpolation by using the sampling grid by the sampler.  

\paragraph{Implementation and Training}
For preprocessing pipeline 1 intramodal registration, the number of iterations had been set as 50, batch size as 2, and the total number of epochs as 2000. For pipeline 2 intramodal registration, the number of iterations was set as 50, total epochs as 1000, and batch size as 2. For intermodal registration with the dataset being preprocessed with pipeline 1, batch size was set to 2, iterations to 100, and epochs equal to 1000. Hyperparameter (taken from the original paper) for the inverse-consistent constraint was set to 0.05, the anti-folding constraint to 100000 and the smoothness constraint to 0.5. The loss was optimised for training using the Adam optimiser with a learning rate of $0.0005$. 

\subsubsection{Deep Learning Baseline: ADMIR}
ADMIR \cite{tang_admiraffine_2020} (Affine and Deformable Medical Image Registration) is an affine and deformable image registration that works end to end. This method does not require the images to be pre-aligned, which in turn helps to do image registration quickly with good accuracy. 

An affine registration network, a deformable registration network, and a spatial transformer are the three fundamental components of ADMIR. Both Affine and Deformable networks were trained at the same time in the original publication. The traditional image registration techniques improve the registration by iteratively optimising the similarity function. 

\paragraph{Architecture and Method Details}
ADMIR \cite{tang_admiraffine_2020} has two sub-networks; one focuses on affine registration, while the other one focuses on deformable registration. The fixed and moving pictures are concatenated and sent into the Affine ConvNet, which predicts 12 affine transformation parameters (rotation, translation, scaling, and shearing) used to calculate the DVF $u_a$, which is then used by the spatial transform to coarsely warp the moving image. The coarsely warped and fixed images are then combined and fed into the Deformable ConvNet, which calculates the DVF $u_d$. To get the final registration result, the final DVF $u_f$ is calculated by aggregating DVF $u_a$ and DVF $u_d$ with the help of a spatial transformer, which warps the moving image to fully register the moving and fixed images. Gradient loss has been employed for smoothness in addition to normalised cross-correlation (NCC) as the similarity loss.

\paragraph{Implementation and Training}
As the focus of the current research is on deformable registration, the affine registration module of ADMIR was not utilised. Only the deformable part of ADMIR was used to perform registration, and consequently, the loss function was adjusted accordingly. The network was trained using the Adam optimiser with a learning rate of $0.0001$ for 1000 epochs. Every image in the training set was successively selected as the moving image and concatenated with the fixed image as a whole to feed into the deformable ConvNet in each epoch.

\subsubsection{Deep Learning Baseline: Voxelmorph}
Voxelmorph \cite{balakrishnan_voxelmorph_2019} is a deep learning based, unsupervised, pairwise, deformable registration framework. The framework also incorporates an additional supervised segmentation training regimen that could help further improve the performance of the dataset. Traditional iterative approaches have the disadvantage of optimising the deformation field for each pair of images independently of others and taking minutes to compute on GPU. Voxelmorph takes in two n-D volumes and computes the deformation field in a few seconds. This reduces computation time whilst maintaining performance - the main motivation for using Voxelmorph as the main inspiration of this current paper.

\paragraph{Architecture and Method Details}
Voxelmorph is based on UNet \cite{iek20163DUL} with LeakyReLU with a parameter \(0.2\) as the activation function that takes a single input of concatenated affinely registered moving and fixed image. In the original work, the scans were resampled to standard \(256 \times 256 \times 256\) with a 1mm isotropic size. The network then predicts the DVF for the supplied pair of fixed-moving images, which is then used to transform moving images using a spatial transformer \cite{Jaderberg2015SpatialTN}. The warped image is then compared against the fixed image using a similarity metric such as MSE and NCC. The deformation field is regularised using a smoothness loss function that penalises unrealistic sharp deformations. Equation \ref{eq:unsup_loss} shows the overall loss function where \(f\) is fixed image, \(m\) is moving image, \(\phi\) is the deformation field and \(\mathcal{L}_{sim}\) is the similarity loss, \(\mathcal{L}_{smooth}\) is the deformation field smoothness loss, \(\lambda\) acts as hyperparameter for regularisation term.
\begin{equation}
    \mathcal{L}_{us}(f, m, \phi) = \mathcal{L}_{sim}(f, m \circ \phi) + \lambda\mathcal{L}_{smooth}(\phi)
    \label{eq:unsup_loss}
\end{equation}

If additional segmentation is considered for training, then the loss function gets modified as below equation \ref{eq:voxelmorph_total_loss} where \(\mathcal{L}_{seg}\) is the segmentation loss

\begin{equation}
    \mathcal{L}_a(f, m, s_f, s_m, \phi) = \mathcal{L}_{us}(f, m, \phi)  + \gamma\mathcal{L}_{seg}(s_f, s_m \circ \phi)
    \label{eq:voxelmorph_total_loss}
\end{equation}

\paragraph{Implementation}
The network was trained with \(\lambda=1\) with a batch size of 5 and was trained for a total of 1000 epochs using the Adam optimiser with a learning rate of $0.0001$. For intramodal registration,  normalised cross-correlation (NCC) and intermodal,  normalised mutual information (NMI) was employed as the loss function.

\subsection{Dataset}
The IXI Dataset~\footnote{IXI Dataset:~\url{https://brain-development.org/ixi-dataset/}} consists of nearly 600 Magnetic resonance (MR) images from healthy subjects. For each subject, the dataset consists of five MR image acquisition protocols, among which T1w and T2w MRIs were used in this research. Intramodal registrations were performed among the different subjects of the T1w images, while the intermodal registrations were performed by taking pairs of T1-T2 MRIs of each subject. The T1 weighted images are with voxel ordering AIL (A-P within I-S within L-R) consisting of 150 slices in each volume, with an in-plane matrix size of 256 x 256 pixels, having a resolution of 0.9375 x 0.9375 x 1.2000\(m^3\)/voxel. The T2 weighted images are with RPI (R-L within P-A within I-S) voxel ordering, comprising 116 to 130 slices in each volume, with the same matrix size and resolution as the T1 weighted images.

\begin{figure}\centering
    \centering
    \includegraphics[width=0.22\textwidth]{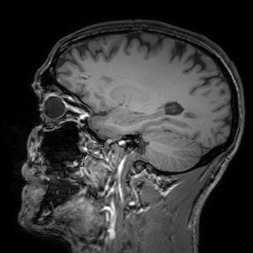}
    \includegraphics[width=0.22\textwidth]{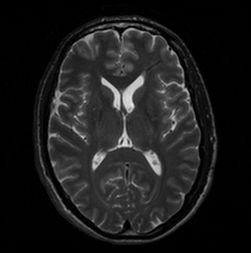}
    \caption{(a) T1-w image (b) T2-w image}
    \label{fig:IXI_Dataset}
\end{figure}

\subsubsection{Data Preprocessing} 
\label{Data_Preprocessing}
Data preprocessing has always been an integral part of machine learning - going hand in hand with the GIGO (Garbage in, Garbage out) concept. These techniques may involve the removal or reduction of noise and artefacts and address the intensity differences of images from different scanners using techniques like image filtering, resampling, and intensity normalisation. 

Preprocessing techniques employed in this research were aimed at 3D brain MRIs. Two different preprocessing pipelines were created - the first pipeline was used as the main pipeline - the models were trained and tested by preprocessing the dataset using this pipeline. The second pipeline was also applied to the test set - to evaluate the generalisation capabilities of the networks while encountering data preprocessed differently than the training set. These pipelines were roughly based on existing research papers with some modifications/adaptations to improve the registration performance. As this research focuses on deformable registration, affine registration was performed as part of the preprocessing steps.

\begin{figure*}
    \centering
    \includegraphics[width=0.8\textwidth]{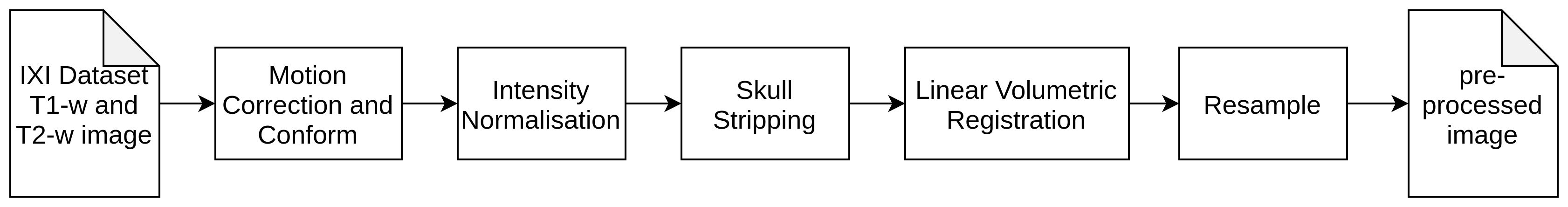}
    \caption{Pre-processing Pipeline 1}
    \label{fig:preprocessing_workflow}
\end{figure*}

\paragraph{Pipeline 1: FreeSurfer Preprocessed Dataset}
\label{sec:pipeline1}
The full preprocessing stream of FreeSurfer~\cite{Dale1999} cortical reconstruction process named \emph{recon-all}, consists of 31 distinct preprocessing stages, divided into sub-directives. Following Voxelmorph~\cite{balakrishnan_voxelmorph_2019}, this research uses the first sub directive, \emph{autorecon1}, which consists of stages as shown in Figure \ref{fig:preprocessing_workflow} and the stages are described below along with the approximate time required for their executions on an AMD Opteron 64bit 2.5GHz processor.

\par \textbf{Motion Correction and Conform:}
The motion correction step takes multiple source volumes as input and corrects for small motions between these volumes, and averages them together. Processing time for the Motion Correction step is estimated to be less than five minutes.

\par \textbf{NU Intensity Correction:}
Variations in scanners or parameters during MR image acquisition often result in intensity non-uniformity in MR data. Hence the non-parametric non-Uniform normalisation (N3) step is incorporated to account for these intensity differences. This step makes relatively few assumptions about the data (non Parametric) and performs four iterations of normalisation. NU intensity correction takes approximately three minutes to process.

\par \textbf{Talairach:}
Talairach is a FreeSurfer script that computes affine transform from the input volume to the MNI305 atlas. The transform computation internally uses the MINC program \emph{mritotal}, and these coordinates are used as seed points in subsequent stages of the cortical reconstruction process. Talairach computation is performed approximately within a minute.

\par \textbf{Normalisation:} 
The normalisation step scales the intensities of all voxels in such a way that the mean intensity of white matter becomes 110, and this step takes approximately three minutes for a volume.

\par \textbf{Skull Strip:} 
Eliminating extra-cranial and non-brain tissues might be an important step to ensure better segmentation of brain regions for many clinical applications and analyses. The skull strip step removes the skull from normalised input volume and generates a brain mask. \emph{mri\_watershed} program is then executed. The whole procedure takes under a minute to complete. The \emph{autorecon1} directive of the FreeSurfer cortical reconstruction process end with this stage.


\par \textbf{Resample:}
The affine registered images were further resampled to a 128 x 128 x 128 grid with 1mm isotropic voxels. The preprocessed dataset was split into 200 image pairs for the training and 50 image pairs for the test. 

The different preprocessing stages with a sample output for each stage are depicted in the Appendix (Figure \ref{fig:preprocessing_stages}).

\paragraph{Pipeline 2}
\label{sec:pipeline2}
\begin{figure*}
    \centering
    \includegraphics[width=0.8\textwidth]{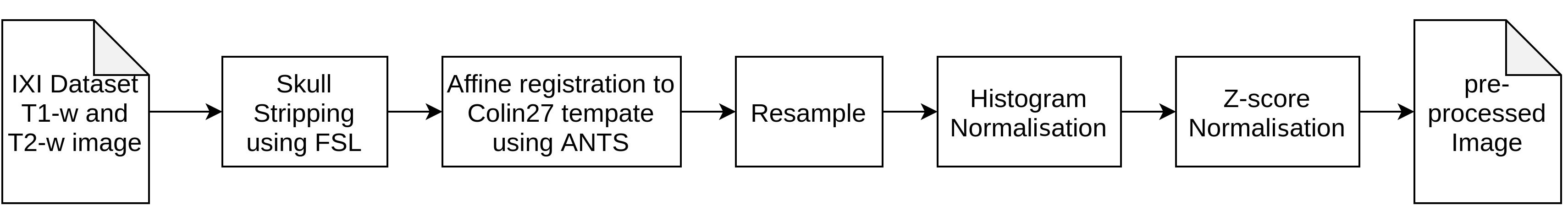}
    \caption{Pre-processing Pipeline 2}
    \label{fig:preprocessing_workflow_icnet}
\end{figure*}
Pipeline 2 follows steps similar to the preprocessing pipeline as performed in ICNet~\cite{zhang_inverse-consistent_2018}. Individual stages of the pipeline are depicted in Figure \ref{fig:preprocessing_workflow_icnet}

\par \textbf{Skull Strip using FSL:} Brain Extraction Tool (BET) of FSL~\citep{smith2002fast,jenkinson2005bet2} is used to extract the brain by removing non-brain tissues from the whole head. 

\par \textbf{Affine registration to Colin27 template:}
The skull-stripped input volumes are then affine registered to the Colin27~\citep{holmes1998enhancement} atlas. Colin27 atlas with high SNR and structure definition is an average of 27 T1 weighted MRI scans of the same individual. This was performed using the Advanced Normalisation Tools (ANTS) toolkit~\citep{avants2009advanced} and took less than a minute to complete.

\par \textbf{Resample:}
After affine registration with Colin27 atlas, the image volumes are of dimensions 181 x 218 x 181. To ensure consistency with the pipeline 1 dataset, they were resampled to a 128 x 128 x 128 grid having isotropic voxels of 1mm. Resampling is performed with the help of nibabel~\citep{brett_matthew_2020_4295521}.

\par \textbf{Histogram and Z-score Normalisation:} Intensity normalisation is performed by matching the intensity histogram of each brain MRI to the Colin27 template using the Histogram matching algorithm. Furthermore, Z-score normalisation was performed to ensure the mean intensity of each image is zero and ensure the standard deviation is one.


\subsection{Evaluation}
\label{Evaluation_metrics}. 
The evaluation of the different methods was performed by comparing the registered images against the fixed images quantitatively and qualitatively. For quantitative evaluation of both intramodal and intermodal registration, the Pearson correlation coefficient, Dice coefficient, and Kullback-Leibler distance were employed. For intramodal registration, Structural Similarity Index and mean-squared error were additionally used. These two metrics were not used for the intermodal registration as they are intensity-based techniques and would not work for evaluation with different modalities. The statistical significance of the changes in all the metric values observed for the different methods was measured using 2-tailed T-tests with a standard significance level of $0.05$. Apart from these quantitative evaluation measures, images were also compared visually.

\par \textbf{Pearson Correlation Coefficient:} Pearson Correlation Coefficient (PCC) is a statistical measure used to understand the departure of two random variables from independence. The Pearson correlation is calculated as 
\begin{equation}
   PCC = 1 / N - 1 \sum_{i=1}^{N} ((x_i - \mu_x)/\sigma_x) ((y_i - \mu_y)/\sigma_y) 
\end{equation}

where \(x_i\) and \(\_i\) are realisations of random variables X and Y and \(\mu_x\) and \(\mu_y\) are the means of X and Y respectively, \(\sigma_x\) and \(\sigma_y\) are standard deviations of $X$ and $Y$ respectively and $N$ is the number of sample pairs.
 
\par \textbf{Dice Coefficient:} Dice coefficient, or simply Dice, is an overlap-based metric extensively used while evaluating image segmentation methods. To compare the registration methods using the Dice score, first, the fixed image and the registered image were segmented into four segments, namely, cerebrospinal fluid (CSF), white matter (WM), grey matter (GM) and the background, using the ANTS atropos segmentation~\citep{avants2011open}. Then the dice score for individual segments was computed and averaged scores to get averaged dice score for the whole volume. Dice score between the fixed image $X$ and registered image $Y$ is computed as: 
\begin{equation}
    DICE = 2|X \cap Y| / |X| + |Y|
\end{equation}
It is to be noted that the segmentation accuracy is a major factor affecting the dice scores as the evaluation relies entirely on the segmented results. 


\par \textbf{Kullback-Leibler Distance:} Kullback-Leibler Distance (KLD) is an evaluation metric used belonging to information theory evaluation measures that give a statistical distance measure between two joint intensity distributions. Given observed and expected intensity distributions as \(p_o(x,y)\) and \(p_s(x,y)\) the KLD is computed as
\begin{equation}
   KLD = \sum_{x\in\chi}\sum_{y\in\chi} p_o(x,y)log(p_o(x,y)/p_e(x,y)) 
\end{equation}


\par \textbf{Structural Similarity Index Measure:} The structural similarity index measure (SSIM)~\citep{wang2004image} is effectively used to evaluate the perceived quality of digital images and the variants of SSIM. The visual image quality is evaluated by measuring the structural similarities between two images where one image is the reference image. Given images $X$ and $Y$ to be compared, and $x$ and $y$ are pairs of square windows of the same size of $X$ and $Y$, the SSIM is calculated as 
\begin{equation}
    SSIM(x,y) = (2\mu_x\mu_y + c_1)(2\sigma_xy + c_2) / (\mu_x^2 + \mu_y^2 + c_1)(\sigma_x^2 + \sigma_y^2 + c_2)
\end{equation}
where \(\mu_x\) and \(\mu_y\) are average pixel values with pixel value standard deviations \(\sigma_x\) and \(\sigma_y\) and co-variance \(\sigma_xy\). The $SSIM(x,y)$ takes a value between 0 and 1, indicating completely different patches to identical patches.

\section{Results and Discussion} \label{results}
The results of the proposed method were compared against the baseline methods quantitatively using the metrics discussed in Sec.~\ref{Evaluation_metrics} and also qualitatively. Initially, different optimisers were compared for direction optimisation (see Sec.~\ref{proposed_method_directoptimization}), and the best-performing optimiser was then used for further comparisons. Comparisons have been performed for the task of intramodal registration (T1w MRIs of different subjects) and intermodal registration (pairwise T1-T2 registration of each subject). 

\subsection{Direct Optimisation}
\begin{figure*}\centering
    \centering
    \subfigure[Fixed volume]{\includegraphics[height=4cm, width=0.25\textwidth]{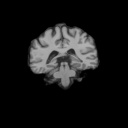}}
    \subfigure[Moving Volume]{\includegraphics[height=4cm, width=0.25\textwidth]{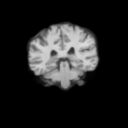}} 
    \subfigure[Adam Optimiser]{\includegraphics[height=4cm, width=0.25\textwidth]{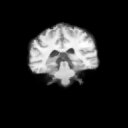}}
    \subfigure[AdamW Optimiser]{\includegraphics[height=4cm, width=0.25\textwidth]{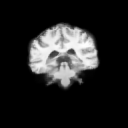}} 
    \subfigure[RMSPROP Optimiser]{\includegraphics[height=4cm, width=0.25\textwidth]{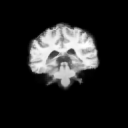}} 
    \subfigure[SGD Optimiser]{\includegraphics[height=4cm, width=0.25\textwidth]{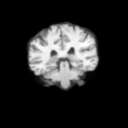}}
    \caption{Intramodal T1 weighted registration using direct optimisation, Coronal view} 
    \label{fig:direct_opt_T1_sag_pipeline1}
\end{figure*}

Four different optimisers were compared using their PyTorch implementations: Adam, AdamW, RMSProp, and SGD. Intramodal and Intermodal results have been reported quantitatively in Table \ref{tab:Intramodal Evaluation for Direct Optimization} and \ref{tab:Intermodal Evaluation for Direct Optimization}. Qualitative results for intermodal registration have been shown in Fig.~\ref{fig:direct_opt_T1_sag_pipeline1}. In three out of five metrics for Intramodal (while the other two favour ADAM) and two out of three for Intermodal, RMSPROP performed better than the other optimisers. Hence, RMSPROP was chosen as the optimiser for direction optimisation for further comparison against the other methods. The optimisation was performed for 1500 epochs and took \(3.47\) minutes each for deformable registration on a 12 GB GPU. 1500 was observed to be the best for the optimisation techniques - both quantitatively and qualitatively. A quantitative comparison of the different number of epochs for the RMSProp optimiser has been shown in the Appendix (Fig.~\ref{tab:Intramodal Direct Optimization Evaluation by Epochs}). Fewer epochs were not enough, while a higher number of epochs resulted in deformation fields that were too smooth or unrealistic. It could mean that constraints should be added to better update the deformation field during the optimisation process. 
\begin{table*}
    \centering
    \small
    \caption{Intramodal Evaluation for Direct Optimisation}
    \label{tab:Intramodal Evaluation for Direct Optimization}
  \begin{tabular}{c c c  c  c c}
    \toprule
    \multirow{2}{*}{Optimiser} &
      \multicolumn{4}{c}{IntraModal} \\
      & {SSIM} & Pearson Correlation & {Dice Score} & MSE & KL Distance \\
      \midrule
    ADAM & \textbf{0.9762 $\pm$ 0.0081} & 0.9886 $\pm$ 0.0041 & 0.7824 $\pm$ 0.0346 & 0.0010 $\pm$ 0.0004 & \textbf{0.0578 $\pm$ 0.0224} \\
    ADAMW &  0.9753 $\pm$ 0.0082 &0.9868 $\pm$ 0.0051 & 0.7836 $\pm$ 0.0355 & 0.0010 $\pm$ 0.0004 & 0.0589 $\pm$ 0.0225 \\
    \textbf{RMSPROP} & 0.9729 $\pm$ 0.0076 & \textbf{0.9912 $\pm$ 0.0017} &  \textbf{0.7926 $\pm$ 0.0195} & \textbf{0.0007 $\pm$ 0.0001} & 0.0649 $\pm$ 0.0183  \\
    SGD & 0.9315 $\pm$ 0.0088 & 0.9584 $\pm$ 0.0073 & 0.6044 $\pm$ 0.0229 & 0.0036 $\pm$ 0.0005 & 0.0665 $\pm$ 0.0223 \\
    \bottomrule
  \end{tabular}
\end{table*}

\begin{table*}
    \centering
    \caption{Intermodal Evaluation for Direct Optimisation}
    \label{tab:Intermodal Evaluation for Direct Optimization}
  \begin{tabular}{c  c   c  c  c}
    \toprule
    \multirow{2}{*}{Optimiser} &
      \multicolumn{3}{c}{InterModal} \\
      & Pearson Correlation & {Dice Score} & KL Distance \\
      \midrule
    ADAM & 0.7695 $\pm$  0.0055 & 0.5908 $\pm$  0.0498 & 0.8780 $\pm$  0.4452 \\
    ADAMW & \textbf{0.8234 $\pm$  0.0344} & 0.6108 $\pm$  0.0524 & 0.8577 $\pm$  0.2838 \\
    \textbf{RMSPROP} & 0.8222 $\pm$  0.0345 & \textbf{0.6114 $\pm$  0.0533} & \textbf{0.8556 $\pm$  0.4435} \\
    SGD & 0.8165 $\pm$  0.0039 & 0.6081 $\pm$  0.0722 & 0.8656 $\pm$  0.4435\\
    \bottomrule
  \end{tabular}
\end{table*}


\subsection{Comparative Results}
The performance of the proposed method was compared quantitatively and qualitatively against three deep learning baseline models: ICNet~\cite{zhang_inverse-consistent_2018},  Voxelmorph~\cite{balakrishnan_voxelmorph_2019}, and ADMIR~\cite{tang_admiraffine_2020}, direct optimisation of the deformation field using the RMSPROP optimiser, and the gold standard - ANTS SyN registration. 

\subsubsection{Intramodel Registration}
\begin{figure*}\centering
    \centering
    \subfigure[Fixed volume]{\includegraphics[height=4cm, width=0.24\textwidth]{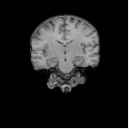}}
    \subfigure[Moving Volume]{\includegraphics[height=4cm, width=0.24\textwidth]{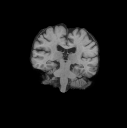}} 
    \subfigure[ANTS SyN registered]{\includegraphics[height=4cm, width=0.24\textwidth]{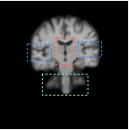}}
    \subfigure[Direct Optimisation registered]{\includegraphics[height=4cm, width=0.24\textwidth]{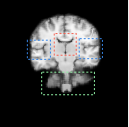}}
     \subfigure[ADMIR registered]{\includegraphics[height=4cm, width=0.24\textwidth]{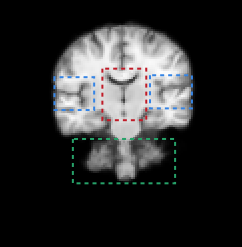}}
    \subfigure[ICNet registered]{\includegraphics[height=4cm, width=0.24\textwidth]{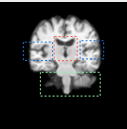}}
    \subfigure[Voxelmorph registered]{\includegraphics[height=4cm, width=0.24\textwidth]{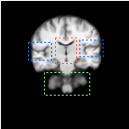}}
    \subfigure[MICDIR registered]{\includegraphics[height=4cm, width=0.24\textwidth]{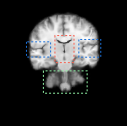}}
    \caption{Intramodal T1 weighted registration Coronal view} 
    \label{fig:intra_registration_T1_coronal}
\end{figure*}

\begin{figure*}\centering
    \centering
    \subfigure[Fixed volume]{\includegraphics[height=4cm, width=0.24\textwidth]{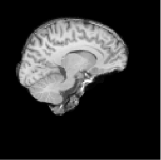}}
    \subfigure[Moving Volume]{\includegraphics[height=4cm, width=0.24\textwidth]{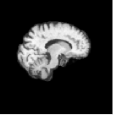}}
    \subfigure[ANTS SyN registered]{\includegraphics[height=4cm, width=0.24\textwidth]{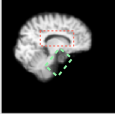}}
    \subfigure[Direct Optimisation registered]{\includegraphics[height=4cm, width=0.24\textwidth]{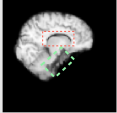}}  
    \subfigure[ADMIR registered]{\includegraphics[height=4cm, width=0.24\textwidth]{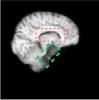}}
    \subfigure[ICNet registered]{\includegraphics[height=4cm, width=0.24\textwidth]{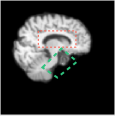}}
    \subfigure[Voxelmoprh registered]{\includegraphics[height=4cm, width=0.24\textwidth]{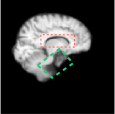}}
    \subfigure[MICDIR registered]{\includegraphics[height=4cm, width=0.24\textwidth]{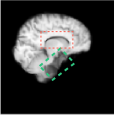}}
    \caption{Intramodal T1 weighted registration Sagittal view} 
    \label{fig:intra_registration_T1_sag}
\end{figure*}

\begin{figure*}\centering
    \centering
    \subfigure[Fixed volume]{\includegraphics[height=4cm, width=0.24\textwidth]{figs/intramodal_pairwise_T1/fixed_sag.png}}
    \subfigure[Moving Volume]{\includegraphics[height=4cm, width=0.24\textwidth]{figs/intramodal_pairwise_T1/moving_sag.png}} 
    \subfigure[ADMIR registered]{\includegraphics[height=4cm, width=0.24\textwidth]{figs/intramodal_pairwise_T1/ADMIR_sag.png}}
    \subfigure[ADMIR DVF]{\includegraphics[height=4cm, width=0.24\textwidth]{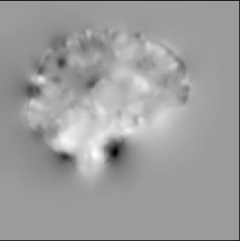}} 
    \subfigure[Voxelmorph registered]{\includegraphics[height=4cm, width=0.24\textwidth]{figs/intramodal_pairwise_T1/vox_sag.png}}
    \subfigure[Voxelmorph DVF]{\includegraphics[height=4cm, width=0.24\textwidth]{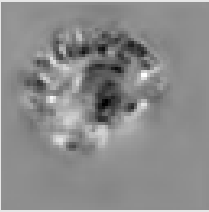}}
    \subfigure[MICDIR registered]{\includegraphics[height=4cm, width=0.24\textwidth]{figs/intramodal_pairwise_T1/scg_sag.png}}
    \subfigure[MICDIR DVF]{\includegraphics[height=4cm, width=0.24\textwidth]{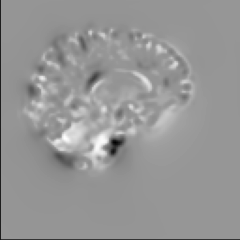}}
    \caption{Intramodal T1 weighted registration Sagittal view} 
    \label{fig:intra_registration_T1_sag_dvf}
\end{figure*}

Table~\ref{tab:intramodal_evaluation_pipeline_1} shows the resulting metric values for the different methods for the task of intramodal registration while being preprocessed with pipeline $1$~(Sec.~\ref{sec:pipeline1}) - the same preprocessing pipeline as the training dataset. The deep learning methods, with the exception of ICNet, consistently seem to outperform the ANTS SyN registration algorithm when evaluated on all the metrics. Among the experimented deep learning baselines, ADMIR was found to be the best-performing one for the intramodal task. Another interesting observation was that the direct optimisation using RMSPROP performed better than the ANTs registration on all metrics while being even better than the ADMIR model on three out of five metrics. The proposed MICDIR outperformed all the baseline models on four out of five metrics with statistical significance while securing the same score as direct optimisation on the last metric - while obtaining statistically significant improvement over the other baselines. The figures \ref{fig:intra_registration_T1_coronal} and \ref{fig:intra_registration_T1_sag} provide visual comparisons of the registration results of the different methods with interesting observable areas of the marked, for coronal and sagittal orientation, respectively. The qualitative comparisons agree with the quantitative evaluation that the proposed MICDIR resulted in better similarities with the fixed image than the baselines. Finally, the deformation vector fields (DVF) generated by the ADMIR and Voxelmoprh are compared qualitatively against the one generated by MICDIR, shown in Fig.~\ref{fig:intra_registration_T1_sag_dvf}. 

\begin{table*}
    \centering
    \small
    \addtolength{\tabcolsep}{-1pt}
    \caption{Intramodal Evaluation on dataset preprocessed with pipeline 1}
    \label{tab:intramodal_evaluation_pipeline_1}
  \begin{tabular}{c  c  c  c  c  c  c}
    \toprule
    \multirow{2}{*}{Algorithm} &
      \multicolumn{5}{c}{IntraModal} \\
      & {SSIM} & PCC & {Dice Score} & MSE & KL Distance \\
      \midrule
    Ants(SyN) & 0.9611 $\pm$  0.0065 & 0.9792 $\pm$  0.0042 & 0.7166 $\pm$  0.0245 & 0.0017 $\pm$  0.0003 & 0.0734 $\pm$  0.0825\\
RMSPROP & 0.9729 $\pm$ 0.0076 & 0.9912 $\pm$ 0.0017 &  0.7926 $\pm$ 0.0195 & \textbf{0.0007 $\pm$ 0.0001} & 0.0649 $\pm$ 0.0183
 \\
    ADMIR & 0.9775 $\pm$  0.0064  & 0.9894 $\pm$  0.0031  & 0.7860 $\pm$  0.0310  & 0.0009 $\pm$  0.0002 & 0.0551 $\pm$  0.0250

 \\
    ICNet & 0.9271 $\pm$  0.0129  & 0.9542 $\pm$  0.0110  & 0.5996 $\pm$  0.0350 & 0.0040 $\pm$  0.0010 & 0.1294 $\pm$  0.2857\\
    VoxelMorph & 0.9711 $\pm$  0.0052 & 0.9878 $\pm$  0.0029 & 0.7747 $\pm$  0.0260 & 0.0010 $\pm$  0.0002 & 0.0535 $\pm$  0.0246
  \\
    \textbf{MICDIR} & \textbf{0.9822 $\pm$  0.0047} & \textbf{0.9916 $\pm$  0.0020} & \textbf{0.8070 $\pm$  0.0234} & \textbf{0.0007 $\pm$  0.0001} & \textbf{0.0506 $\pm$  0.0222}
 \\
    \bottomrule
  \end{tabular}
\end{table*}

\begin{figure*}\centering
    \centering
    \subfigure[Fixed volume]{\includegraphics[height=4cm, width=0.25\textwidth]{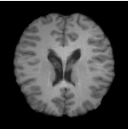}}
    \subfigure[Moving Volume]{\includegraphics[height=4cm, width=0.25\textwidth]{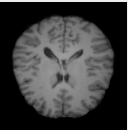}} 
    \subfigure[ANTS SyN registered]{\includegraphics[height=4cm, width=0.25\textwidth]{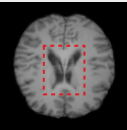}}
    \subfigure[ICNet registered]{\includegraphics[height=4cm, width=0.25\textwidth]{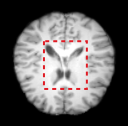}} 
    \subfigure[Voxelmoprh registered]{\includegraphics[height=4cm, width=0.25\textwidth]{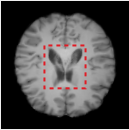}} 
    \subfigure[MICDIR registered]{\includegraphics[height=4cm, width=0.25\textwidth]{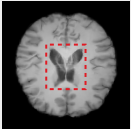}}
    \caption{Intramodal T1 weighted registration with data preprocessed using pipeline 2, Axial view} 
    \label{fig:intra_registration_T1_sag_pipeline2}
\end{figure*}

\begin{figure*}\centering
    \centering
    \subfigure[Fixed Volume]{\includegraphics[height=4cm, width=0.24\textwidth]{figs/intramodal_template_T1/fixed.png}}
    \subfigure[Moving Volume]{\includegraphics[height=4cm, width=0.24\textwidth]{figs/intramodal_template_T1/moving.png}} 
    \subfigure[ICNet registered]{\includegraphics[height=4cm, width=0.24\textwidth]{figs/intramodal_template_T1/icnet.png}}
    \subfigure[ICNet DVF]{\includegraphics[height=4cm, width=0.24\textwidth]{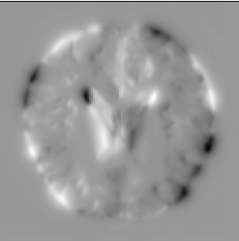}} 
    \subfigure[Voxelmorph registered]{\includegraphics[height=4cm, width=0.24\textwidth]{figs/intramodal_template_T1/vox.png}}
    \subfigure[Voxelmorph DVF]{\includegraphics[height=4cm, width=0.24\textwidth]{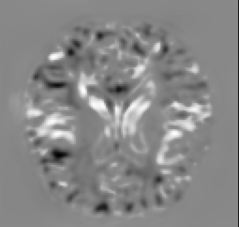}}
    \subfigure[MICDIR registered]{\includegraphics[height=4cm, width=0.24\textwidth]{figs/intramodal_template_T1/scg.png}}
    \subfigure[MICDIR DVF]{\includegraphics[height=4cm, width=0.24\textwidth]{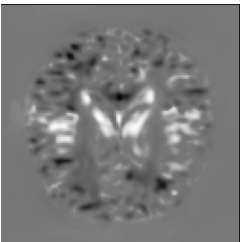}} 
    \caption{Intramodal T1 weighted registration on pipeline 2 Axial view} 
    \label{fig:intra_registration_T1_axial_dvf}
\end{figure*}

The performance of these networks was further evaluated on data preprocessed with pipeline $2$ while being trained on data preprocessed with pipeline $1$ - to compare the generalisation performance of these networks on data preprocessed differently than training. As observed from table \ref{tab:intramodal_evaluation_pipeline_2}, the MICDIR showed significant improvement in all four metrics when compared to gold standards ANTS SyN and deep learning networks ICNet and Voxelmorph. These results strongly suggest that the MICDIR can also generalise much better than the other deep learning networks and provides better performance even with respect to the ANTS SyN registration algorithm, even when trained with different preprocessing techniques. Figures~\ref{fig:intra_registration_T1_sag_pipeline2}~and~\ref{fig:intra_registration_T1_axial_dvf} show the superiority of the proposed MICDIR over the baselines qualitatively in terms of the registration quality and the DVFs, respectively.

\begin{table*}
    \centering
    \caption{Intramodal Evaluation on dataset preprocessed with pipeline 2}
    \label{tab:intramodal_evaluation_pipeline_2}
  \begin{tabular}{c  c  c  c  c c}
    \toprule
    \multirow{1}{*}{Algorithm} &
      \multicolumn{4}{c}{IntraModal} \\
      & {SSIM} & PCC & {Dice Score}  & {MSE} \\
      \midrule
    Ants(SyN) & 0.8734 $\pm$  0.0326 & 0.9702 $\pm$  0.0189 &  0.7103 $\pm$  0.0616 & 0.0051 $\pm$  0.0037\\
    ICNet & 0.7879 $\pm$  0.0576 & 0.9371 $\pm$  0.0243  & 0.5598 $\pm$  0.1181 & 0.0098 $\pm$  0.0062\\
    VoxelMorph & 0.8851 $\pm$  0.0260 &  0.9724 $\pm$  0.0130 & 0.6999 $\pm$  0.0580 & 0.0048 $\pm$  0.0030\\
    \textbf{MICDIR} & \textbf{0.9025 $\pm$  0.0236} & \textbf{0.9761 $\pm$  0.0116}  & \textbf{0.7445 $\pm$  0.0447} & \textbf{0.0041 $\pm$  0.0037}\\
    \bottomrule
  \end{tabular}
\end{table*}

\begin{figure*}\centering
    \centering
    \subfigure[Fixed volume]{\includegraphics[height=5cm, width=0.32\textwidth]{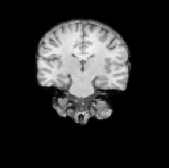}}
    \subfigure[Moving Volume (Colin27)]{\includegraphics[height=5cm, width=0.32\textwidth]{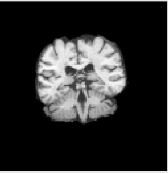}}
    \subfigure[ANTS SyN registered]{\includegraphics[height=5cm, width=0.32\textwidth]{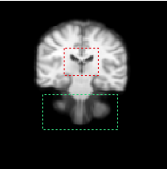}}
    \subfigure[ADMIR registered]{\includegraphics[height=5cm, width=0.32\textwidth]{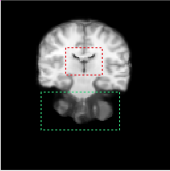}}  
    \subfigure[Voxelmorph registered]{\includegraphics[height=5cm, width=0.32\textwidth]{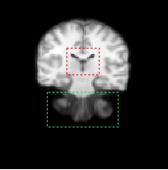}}
    \subfigure[MICDIR registered]{\includegraphics[height=5cm, width=0.32\textwidth]{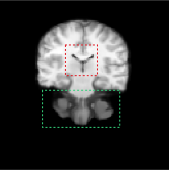}}
    \caption{Intramodal T1 weighted registration with Colin27 template Coronal view} 
    \label{fig:intra_registration_T1_colin_coronal}
\end{figure*}

\paragraph{Template Registration:}
One use-case scenario for intramodal registration is to register a template image as the moving image to a given target image. After obtaining the DVF, it can then be applied to the segmentation mask of the template image to transform it into the segmentation mask of the target image - to perform template-based segmentations. This use-case was evaluated by using the Colin27 template~\citep{holmes1998enhancement} as the moving image and registering it against the images from the test set as fixed images. The images were processed using preprocessing pipeline 1. Table~\ref{tab:intramodal_evaluation_pipeline_1_colin} presents the resultant metric values, and the proposed method outperformed all the baseline methods in four out of five metrics (with statistical significance?). Fig.~\ref{fig:intra_registration_T1_colin_coronal} shows an example registration for visual comparison, where the qualitative results corroborate the results of the quantitative evaluation.

\begin{table*}
    \centering
    \small
    \addtolength{\tabcolsep}{-1pt}
    \caption{Intramodal Evaluation on dataset preprocessed with pipeline 1 registered to Colin27 template}
    \label{tab:intramodal_evaluation_pipeline_1_colin}
  \begin{tabular}{c  c  c  c  c  c  c}
    \toprule
    \multirow{2}{*}{Algorithm} &
      \multicolumn{5}{c}{IntraModal} \\
      & {SSIM} & PCC & {Dice Score} & MSE & KL Distance \\
      \midrule
    Ants(SyN) & 0.9641 $\pm$ 0.0058 & 0.9818 $\pm$ 0.0028 &  0.7420 $\pm$ 0.0178 & 0.0015 $\pm$ 0.0002 & 0.0567 $\pm$ 0.0184\\
    RMSPROP & 0.9774 $\pm$ 0.0048 & 0.9895 $\pm$ 0.0020  & 0.7957 $\pm$ 0.0138 & 0.0009 $\pm$ 0.0002 &  \textbf{0.0445 $\pm$ 0.0121}\\ 
    ADMIR & 0.9803 $\pm$ 0.0034 & 0.9909 $\pm$ 0.0017  & 0.8071 $\pm$ 0.0147 & 0.0007 $\pm$ 0.0001 & 0.0496 $\pm$ 0.0167 \\ 
    VoxelMorph & 0.9779 $\pm$ 0.0042 &  0.9885 $\pm$ 0.0023 & 0.7901 $\pm$ 0.0159 & 0.0009 $\pm$ 0.0002 & 0.0463 $\pm$ 0.0192 \\
    \textbf{MICDIR} & \textbf{0.9840 $\pm$  0.0030} & \textbf{0.9923 $\pm$  0.0014} & \textbf{0.8221 $\pm$  0.0134} & \textbf{0.0006 $\pm$  0.0001} & 0.0448 $\pm$  0.0173
 \\
    \bottomrule
  \end{tabular}
\end{table*}

\begin{figure*}
    \centering
    \subfigure[Fixed volume]{\includegraphics[height=4cm, width=0.24\textwidth]{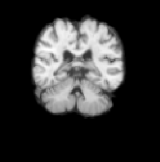}}
    \subfigure[Moving Volume]{\includegraphics[height=4cm, width=0.24\textwidth]{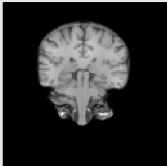}} 
    \subfigure[ANTS SyN registered]{\includegraphics[height=4cm, width=0.24\textwidth]{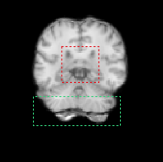}}
    \subfigure[Direct Optimisation registered]{\includegraphics[height=4cm, width=0.24\textwidth]{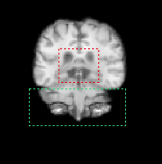}}
     \subfigure[ADMIR registered]{\includegraphics[height=4cm, width=0.24\textwidth]{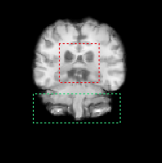}}
    \subfigure[Voxelmorph registered]{\includegraphics[height=4cm, width=0.24\textwidth]{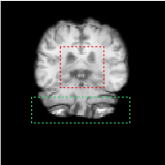}}
    \subfigure[MICDIR registered]{\includegraphics[height=4cm, width=0.24\textwidth]{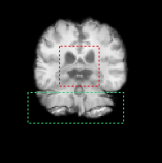}}
    \caption{Intramodal T1 weighted registration with fixed and moving images flipped (original moving as the new fixed and the original fixed as the new moving) Coronal view} 
    \label{fig:intra_registration_T1_coronal_flipped}
\end{figure*}

\paragraph{Inverse-Consistency:}
MICDIR employs the cycle consistency loss, which makes the MSCGUNet inverse-consistent - making it possible for the network to register in both directions: moving to fixed - the intended direction, also used during training all the models (the proposed, as well as the baseline models), as well as fixed to moving. Evaluations were performed to assess the inverse consistency of the model. For these evaluations, MICDIR and the baselines received the fixed as moving and moving as fixed. It is to be noted that the deep learning models (MICDIR, as well as the baselines) were trained to register moving images to the fixed images, while the baseline deep learning models were not inverse-consistent and might not be a fair comparison as without inverse-consistency, this swap of fixed-moving images might be difficult to handle for those models. However, the non-deep learning baselines are unaffected by this change as they do not involve any training. In this scenario as well, MICDIR demonstrated statistically significant improvements over all the baselines; quantitative results are presented in Table~\ref{tab:intramodal_evaluation_pipeline_1_inverse_consistency}, while the qualitative comparison has been shown in Fig.~\ref{fig:intra_registration_T1_coronal_flipped}. Statistical tests were also performed for each method comparing the registration quality of moving to fixed (original registration order, results are in Table~\ref{tab:intramodal_evaluation_pipeline_1}) and fixed to moving (flipped testing, presented in Table~\ref{tab:intramodal_evaluation_pipeline_1_inverse_consistency}). The proposed MICDIR, as well as the non-deep learning baselines (ANTs and RMSPROP), showed no statistically significant difference, while the differences in the metric values of the deep learning baselines (ADMIR and VoxelMorph) were statistically significant. This evaluation conclusively demonstrates the inverse consistency of MICDIR. 

\begin{table*}
    \centering
    \small
    \addtolength{\tabcolsep}{-1pt}
    \caption{Intramodal Evaluation on dataset preprocessed with pipeline 1 with fixed and moving images flipped}
    \label{tab:intramodal_evaluation_pipeline_1_inverse_consistency}
  \begin{tabular}{c  c  c  c  c  c  c}
    \toprule
    \multirow{2}{*}{Algorithm} &
      \multicolumn{5}{c}{IntraModal} \\
      & {SSIM} & PCC & {Dice Score} & MSE & KL Distance \\
      \midrule
    Ants(SyN) & 0.9613 $\pm$ 0.0071 & 0.9805 $\pm$ 0.0045 &  0.7361 $\pm$ 0.0247 & 0.0018 $\pm$ 0.0006 & 0.0486 $\pm$ 0.0920\\
    RMSPROP & 0.9740 $\pm$ 0.0087 & 0.9876 $\pm$ 0.0050  & 0.7935 $\pm$ 0.0315 & 0.0012 $\pm$ 0.0007 &  0.0417 $\pm$ 0.0128\\ 
    ADMIR & 0.9664 $\pm$ 0.0074 & 0.9796 $\pm$ 0.0041  & 0.7822 $\pm$ 0.0289 & 0.0010 $\pm$ 0.0005 & 0.0360 $\pm$ 0.0744 \\ 
    VoxelMorph & 0.9637 $\pm$ 0.0076 &  0.9771 $\pm$ 0.0047 & 0.7670 $\pm$ 0.0313 & 0.0012 $\pm$ 0.0005 & 0.0420 $\pm$ 0.0806 \\
    \textbf{MICDIR} & \textbf{0.9825 $\pm$  0.0054} & \textbf{0.9922 $\pm$  0.0024} & \textbf{0.8101 $\pm$  0.0212} & \textbf{0.0008 $\pm$  0.0005} & \textbf{0.0341 $\pm$  0.0093}
 \\
    \bottomrule
  \end{tabular}
\end{table*}

\subsubsection{Intermodal Registration}
Experiments were conducted registering volumes of different modalities, namely T1 weighted and T2 weighted volumes, pairwise for each subject. The intermodal registration was performed, keeping T1-weighted volumes as fixed volumes and T2-weighted volumes as moving volumes. SSIM and mean squared error (MSE) evaluation metrics were not used for evaluating intermodal registrations as SSIM and MSE are intensity-based metrics and will fail to provide satisfactory results for intermodal cases. As described by \cite{Razlighi2013}, Kullback-Leibler Distance (KLD) which belongs to the class of information-theoretic measures, was used, and Pearson Correlation coefficient, which is a statistical evaluation measure as described in detail in section \ref{Evaluation_metrics}.

\begin{table}
    \centering
    \caption{Intermodal Evaluation}
    \label{tab:intermodal_evaluation_pipeline_1}
    \resizebox{0.48\textwidth}{!}{%
  \begin{tabular}{c  c  c  c c}
    \toprule
    \multirow{2}{*}{Algorithm} &
      \multicolumn{3}{c}{InterModal} \\
      & Pearson Correlation & {Dice Score} & KL Distance\\
      \midrule
    Ants(SyN) & \textbf{0.8555 $\pm$  0.0263}  & 0.5243 $\pm$  0.1318 & 0.8690 $\pm$  0.4904 \\
    ICNet   &  0.7490 $\pm$  0.0640 & 0.5921 $\pm$  0.0528 & 0.9398 $\pm$  0.4124  \\
    VoxelMorph  & 0.8525 $\pm$  0.0245  & 0.6071 $\pm$  0.0510 & \textbf{0.8290 $\pm$  0.4575} \\
    \textbf{MICDIR} & 0.8539 $\pm$  0.0245  & \textbf{0.6211 $\pm$  0.0309} & 0.8338 $\pm$  0.4587 \\
    \bottomrule
  \end{tabular}}
\end{table}

\begin{figure*}\centering
    \centering
    \subfigure[Fixed volume]{\includegraphics[height=4.5cm, width=0.27\textwidth]{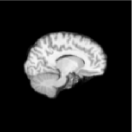}}
    \subfigure[Moving Volume]{\includegraphics[height=4.5cm, width=0.27\textwidth]{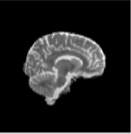}}
    \subfigure[ANTS SyN registered]{\includegraphics[height=4.5cm, width=0.27\textwidth]{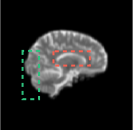}}
    \subfigure[ICNet registered]{\includegraphics[height=4.5cm, width=0.27\textwidth]{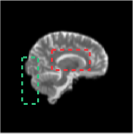}} 
    \subfigure[Voxelmoprh registered]{\includegraphics[height=4.5cm, width=0.27\textwidth]{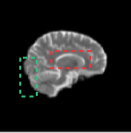}} 
    \subfigure[MICDIR registered]{\includegraphics[height=4.5cm, width=0.27\textwidth]{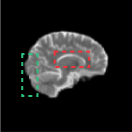}}
    \caption{Intermodal registration with data preprocessed using pipeline 2, Sagittal view} 
    \label{fig:inter_registration_T1_sag_pipeline1}
\end{figure*}

As observed from table \ref{tab:intermodal_evaluation_pipeline_1}, the MICDIR performs better than ANTS SyN, ICNet and Voxelmorph, considering the Dice score and KL distance. Although the MICDIR performs better than the other baselines, the improvement was not statistically significant. In contrast, the ANTs registration resulted in the highest Pearson correlation value, while MICDIR is the second best - but the difference is statistically insignificant. Visual comparisons in Fig.~\ref{fig:inter_registration_T1_sag_pipeline1} reveal that the CSF (red box) is mapped slightly better in the MICDIR compared to other deep learning networks while also able to register the overall brain structure, including the edges (green box), better than the other methods. 

\subsection{Ablation Study of MICDIR}
\begin{table}
\centering
 \caption{Intramodal registration performance evaluation for Test set preprocessed with pipeline 1 T1 weighted volumes for different network configurations }
    \label{tab:proposed_network_config}
\begin{tabular}{c c c c c}
\toprule
\multicolumn{3}{c}{Configuration} & \multicolumn{2}{c}{Evaluation Metric}  \\
SCG & MSS & IC                 & SSIM & DICE                            \\
\midrule
x & x & x                & 0.9713 $\pm$  0.0048 & 0.7738 $\pm$  0.0243                             \\
x & x & \checkmark                & 0.9767 $\pm$  0.0047 & 0.7866 $\pm$  0.0228                          \\
x & \checkmark & x               & 0.9782 $\pm$  0.0053 & 0.7902 $\pm$  0.0211                            \\
\checkmark & x & x               & 0.9757 $\pm$  0.0051 & 0.7760 $\pm$  0.0232                      \\
x & \checkmark & \checkmark               & 0.9802 $\pm$  0.0047 & 0.7980 $\pm$  0.0233                             \\
\checkmark & \checkmark & \checkmark               & \textbf{0.9822 $\pm$  0.0047} & \textbf{0.8070 $\pm$  0.0234}      \\
\bottomrule
\end{tabular}
\end{table}

As described in detail in section \ref{proposed_method_mscgunet}, MICDIR extends the UNET architecture by adding the concepts of encoding global dependencies by employing self-constructing graph network (SCG), multi-scale supervision for faster convergence (MSS) and better attention to detail in different scales, and finally cycle consistency (IC) - which ensures pair of images are symmetrically deformed towards one another. Hence, it is interesting to evaluate the contribution of the different modifications. An ablation study was conducted by adding each of the modifications separately and observing the changes in the performance in terms of SSIM and Dice. The models were trained and tested on data preprocessed with pipeline 1. It can be observed that each of the modifications (SCG, MSS, IC) improved the performance of the base model - while the configurations involving MSS and IC significantly improved the registration. A combination of MSS and ICC further improved the performance of both models with MSS and IC, while the final MICDIR comprising all three modifications (SCG, MSS, IC) resulted in the best-performing model. 

\begin{table}
\centering
 \caption{Intramodal registration performance evaluation for Test set preprocessed with pipeline 2 T1 weighted volumes with and without SCG in latent }
    \label{tab:proposed_network_config_pipeline2}
\begin{tabular}{c c c c c}
\toprule
\multicolumn{3}{c}{Configuration} & \multicolumn{2}{c}{Evaluation Metric}  \\
SCG & MSS & IC                 & SSIM & DICE                            \\
\midrule
x & x & x                & 0.8794 $\pm$  0.0495 & 0.6768 $\pm$  0.1305                             \\
\checkmark & x & x       & \textbf{0.8875 $\pm$  0.0504} & \textbf{0.6905 $\pm$  0.1335}                             \\
\bottomrule
\end{tabular}
\end{table}

From this ablation study, it was observed that the SCG component marginally improves the scores when compared to the base network. So an additional experiment was performed by comparing the base network and the base network with SCG by testing on data preprocessed with pipeline 2. In this experiment, a significant improvement can be observed with SCG in terms of both SSIM and Dice. Hence, it can be concluded that the SCG component helps improve the network's generalisation capabilities.

\section{Conclusions and Future Work}
\label{sec:conclusions}
This paper proposed MICDIR - multi-scale inverse-consistent deformable image registration using the novel MSCGUNet - a multi-scale UNet with a self-constructing graph at its latent space and showed its application for the task of intramodal and intermodal image registration. The proposed network allows to explicitly encode global dependencies and semantics present in the given image pairs, i.e. structure and overall view of the anatomy in the supplied image, by incorporating a self-constructing graph network in the latent space of a UNet model. The multi-scale architecture helps track different amounts of deformations, and the inverse consistency constraint ensures that the deformations are consistent. The proposed method was compared against three deep learning based deformable registration techniques, direct optimisation baseline - employing optimisers used in deep learning to optimise the deformation field directly, and against a gold standard registration pipeline. For the task of intramodal registration, where T1w MR volumes of different subjects were co-registered to the same coordinate space, the proposed method outperformed all the baseline models, and the improvements observed were statistically significant. For intermodal registration, where pairwise registrations were performed for different subjects, co-registering T2w images to T1w images, the proposed method also performed better than the baselines. Furthermore, an ablation study of the proposed method was performed, which shows how much the different modifications are helping in improving the performance. Experiments have revealed that the MICDIR not only outperforms the baselines but also generalises better - making it more robust against variations in the input. 

The performance of the proposed method might further be improved by designing a loss function that targets transformer loss - just like the SCG loss, and also by integrating an affine network preceding the deformable network and training them end-to-end. The deformation field could be optimised better by adding other losses, such as anti-folding loss. The performance of the intermodal registration could be improved by converting it to an intramodal registration problem - by harmonising them into a pseudo-similar modality~\cite{Zuo2021InformationbasedDR} and then registering it using the proposed MICDIR as an intramodal problem and then training it end-to-end.

\section*{Acknowledgement}
This work was in part conducted within the context of the International Graduate School MEMoRIAL at Otto von Guericke
University (OVGU) Magdeburg, Germany, kindly supported by the European Structural and Investment Funds (ESF) under the
programme "Sachsen-Anhalt WISSENSCHAFT Internationalisierung" (project no. ZS/2016/08/80646).

\bibliography{mybibfile}

\appendix
\begin{figure*}[htb!]\centering
    \centering
    \includegraphics[width=0.25\textwidth]{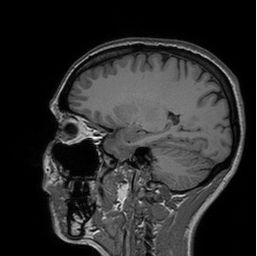} 
    \includegraphics[width=0.25\textwidth]{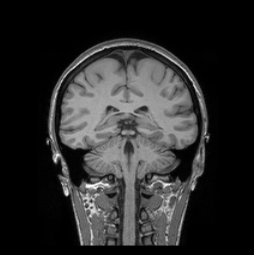} \includegraphics[width=0.25\textwidth]{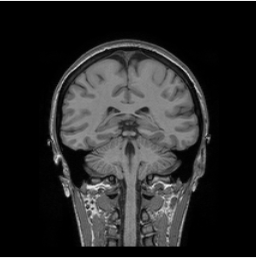}
    \includegraphics[width=0.25\textwidth]{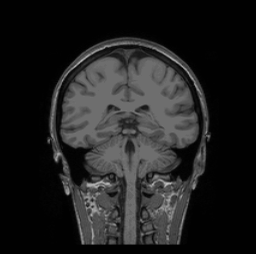}
    \includegraphics[width=0.25\textwidth]{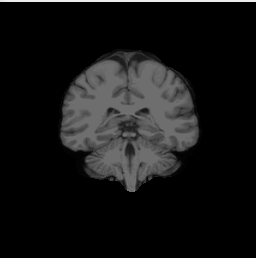}
    \includegraphics[width=0.25\textwidth]{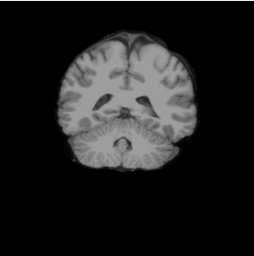}
    \caption{(a) Input Volume (256 x 256 x 150) 0.9375 x 0.9375 x 1.2000\(m^3\)/voxel (b)Motion Correct and Conform (256 x 256 x 256) 1mm isotropic voxels (c) Intensity Correction for better segmentation (d) Intensity Normalised (e) Skull Stripped volume (f) Affine registered volume }
    \label{fig:preprocessing_stages}
\end{figure*}


\begin{table*}[htb!]
    \centering
    \small
    \caption{Intramodal Direct Optimisation Evaluation by Epochs}
    \label{tab:Intramodal Direct Optimization Evaluation by Epochs}
  \begin{tabular}{c c c  c  c c}
    \toprule
    \multirow{2}{*}{Epochs} &
      \multicolumn{4}{c}{IntraModal - RMSPROP} \\
      & {SSIM} & Pearson Correlation & {Dice Score} & MSE \\
      \midrule
    1000 & $0.9713\pm0.0042$ & $0.9831\pm0.0023$ & $0.7901\pm0.0293$ & $0.0010\pm0.0002$ \\
    1500 & $0.9729\pm0.0076$ & $0.9912\pm0.0017$ & $0.7926\pm0.0195$ & $0.0007\pm0.0001$  \\
    2000 & $0.9725\pm0.0063$ & $0.9910\pm0.0025$ & $0.7919\pm0.0271$ & $0.0008\pm0.0002$ \\
    2500 & $0.9722\pm0.0069$ & $0.9871\pm0.0022$ & $0.7881\pm0.0313$ & $0.0009\pm0.0002$ \\
    \bottomrule
  \end{tabular}
\end{table*}

\end{document}